# A multi-channel DAQ system based on FPGA for long-distance transmission in nuclear physics experiments

Hongwei Yu, Kezhu Song, Junfeng Yang, Kehan Li, Tengfei Chen, Shiyu Luo, Cheng Tang, Han Yu

*Abstract*—As the development of electronic science and technology, electronic data acquisition (DAQ) system is more and more widely applied to nuclear physics experiments. Workstations are often utilized for data storage, data display, data processing and data analysis by researchers. Nevertheless, the workstations are ordinarily separated from detectors in nuclear physics experiments by several kilometers or even tens of kilometers. Thus a DAQ system that can transmit data for long distance is in demand. In this paper, we designed a DAQ system suitable for high-speed and high-precision sampling for remote data transfer. An 8-channel, 24-bit simultaneous sampling analog-to-digital converter(ADC) named AD7779 was utilized for high-speed and high-precision sampling, the maximum operating speed of which runs up to 16 kilo samples per second(KSPS). ADC is responsible for collecting signals from detectors, which is sent to Field Programmable Gate Array(FPGA) for processing and long-distance transmission to the workstation through optical fiber. As the central processing unit of DAQ system, FPGA provides powerful computing capability and has enough flexibility. The most prominent feature of the system is real-time mass data transfer based on streaming transmission mode, highly reliable data transmission based on error detection and correction and high-speed high-precision data acquisition. The results of our tests show that the system is able to transmit data stably at the bandwidth of 1Gbps.

*Index Terms*—Data acquisition, high-speed and high precision sampling, long-distance transmission

## I. Introduction

Data acquisition(DAQ) system is an important information system in nuclear physics experiment. The main function of DAQ system is to collect signal from detectors in physics experiment and transmit signal to workstations through transfer medium. The workstations receive and store data in order that researchers complete data processing and analyzing to get the information they want.

With development of science and technology, the scale of physics experiments is getting larger and larger. Large-scale physics experiments are usually accompanied by several thousand channels of data, the order of magnitude of which reaches Gbps. Meanwhile the detectors and workstations are often separated by several kilometers or even tens of kilometers. Therefore a DAQ system for large capacity and remote real time transmission is needed. Optical fiber transmission is one of the most common method of long-distance data transfer. The advantages of optical fiber transmission are high bandwidth, no interruption from electromagnetic noise and long-distance transmission. Generally data transfer speed of single-mode optical fiber reaches Gb/s for several kilometers[1]. Considering the advantages of optical fiber transmission, we designed a DAQ system with large-capacity and remote real time data transmission based on streaming mode, the core of which is FPGA. Taking advantage of hardware parallelism, FPGAs provide more powerful computing ability than digital signal processors (DSPs) and offer more flexibility. The simplified prototype architecture of nuclear physics experiments is shown in Fig.1.

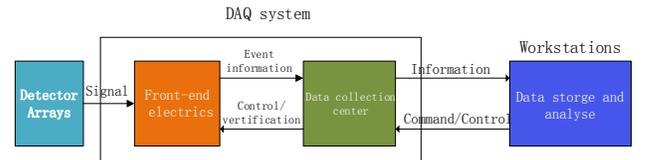

Fig.1 Simplified prototype architecture of physics experiments

One of the advantages of the system is real-time transfer of large-capacity data based on streaming transmission mode , which utilizes error detection and correction mechanism to ensure the transmission of high reliability. Another major advantage of the system is high-speed high-precision data acquisition and remote transmission by optical fiber. The system utilizes an 8-channel, 24-bit simultaneous sampling analog-to-digital converter (ADC) at a speed of 16 kilo samples per second (KSPS)[4]. Signal-to-noise ratio(SNR) reaches 108dB at 16 KSPS in high resolution mode.

[1]This work was supported by National Key R&D Plan of the Ministry of Science and Technology of China under Grant No. 2016YFC0303200.
[2]The authors are with the State Key Laboratory of Particle Detection and Electronics, University of Science and Technology of China, Hefei 230026, China and also with Department of Modern Physics, University of Science and Technology of China, Hefei 230026 , China(email: yhw1993@mail.ustc.edu.cn;skz@ustc.edu.cn;yangjf@ustc.edu.cn).

## II. THE DAQ SYSTEM ARCHITECTURE

### A. Hardware

According to the design requirements, the functional block diagram of core circuit board of DAQ system is shown as Fig.2. The circuit board consists of the on-board power system, FPGA, optical interface unit, electrical interface unit and ADC module. We utilized the Xilinx Kintex-7 FPGA as the central processing unit so as to guarantee that the FPGA have sufficient memory blocks and process data with adequate speed[5]. The FPGA is configured by a Serial Peripheral Interface(SPI) flash and the system clock is 40MHz provided by external crystal oscillator[3]. For electrical interface, low-voltage differential signaling (LVDS) technology is used to exchange data and command. LVDS is a current-driven differential signal transmission technology. It is suitable for high-speed transmission with short-distance. The data transmission rate of LVDS is up to 655Mbps, which meets most of the data transfer requirements of physics experiments. The key elements of optical interface is Small Form-factor Pluggable(SFP) transceiver, which is a small, hot-pluggable optical transceivers for optical communications applications. A kind of SFP transceiver was applied to exchange data and command between the circuit board and workstation, the transmitting rate of which reaches 1Gbps and is far exceeding the system bandwidth of the prototype architecture.

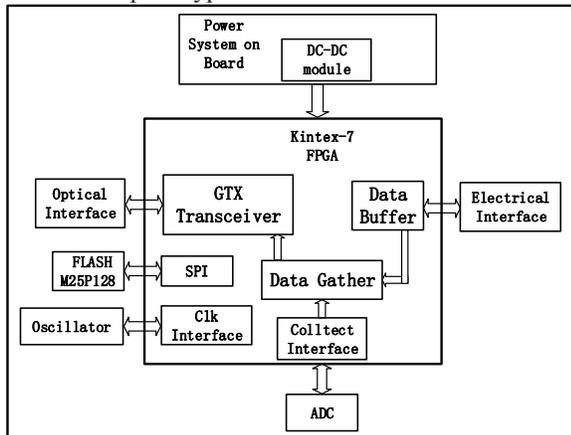

Fig.2 Simplified architecture of the core circuit board

### B. Signal Flow

The simplified signal flow of core circuit board is shown in Fig.3. In physics experiments, ADC acquires analog signal from detector and converts it to digital signal. FPGA receives signal from ADC and sends command signal to ADC. At the same time, electrical interface is used to receive data from other circuit boards. The two part of data is merged in First Input First Output(FIFO) inside of FPGA and then sent to workstation for storage and analysis. The core board mainly undertakes collection of local data and transmission of external data and local data.

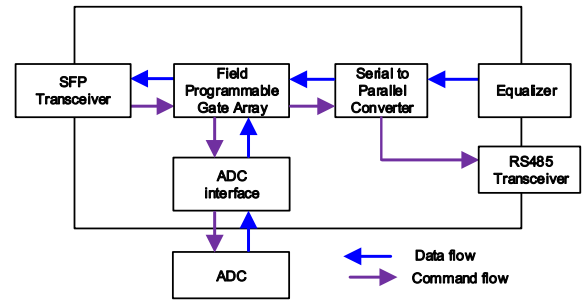

Fig.3. Signal Flow

## III. STREAMING TRANSMISSION MODE BASED ON FPGA

The kernel of the system is the flow of data and command streams between different FPGAs. Based on the open system interconnection(OSI), we designed a transmission module as one of the core of streaming transmission mode. As shown in Fig.4, the transmission module is used to implement point-to-point communication, which corresponds to the physical layer and data link layer in the OSI model.

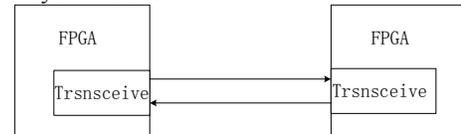

Fig.4 Simplified Transmission module

The transmission module can be divided into four independent modules according to functions and tasks, namely PHY, SYN, MAC and LLC. As shown in Fig.5, PHY module corresponds to the physical layer in the OSI model, which is used to achieve 8B / 10B encoding, clock data recovery(CDR), serial-to-parallel conversion and other functions. The SYN module handles all synchronize data to ensure timeliness of synchronization commands. MAC module achieves the data frame encapsulation. LLC module realizes data segmented and reorganization, sequential transmission and error retransmission. All four modules use standard interface.

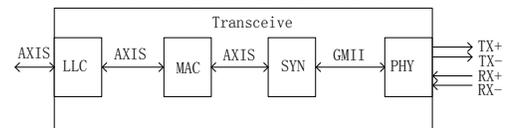

Fig.5 Four parts of Transmission module

Internal composition of PHY module is diverse in terms of different transmission media. We choose optical fiber to transmit data from the system to the workstation. As for optical fiber transmission, PHY module is achieved by FPGA of Xilinx company , which internally integrates GTX transceiver[2]. GTX transceiver realizes high-speed serial to parallel conversion, clock data recovery, 8B/10B encoding and decoding functions. For low-speed signal transmission inside of system, we design a PHY module shown as Fig.6. This module is capable of realizing most functions as GTX transceiver does. Besides Cyclic Redundancy Check( CRC) checking mechanism is applied to judge whether errors appear in the process of  transmission , which is a part of ensuring high reliable transmission.

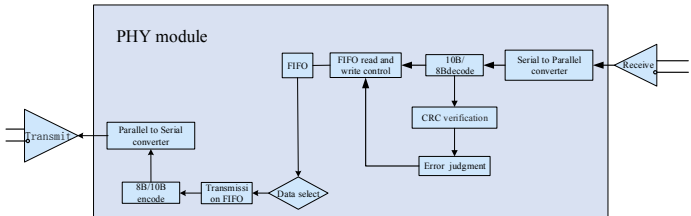
Fig.6. PHY module architecture

## IV. TEST RESULT

The core board of the system is mainly responsible for data summary and transmitting data to workstations remotely. To transmit data to workstation over long distance, we have used the GTX transceiver of Kintex-7 FPGA[2]. It is the physical layer foundation that achieves high-speed serial data transmission, of which the line speed reaches up to 500Mb/s-12.5Gb/s. Meanwhile, the Integrated Bit Error Ratio Tester(IBERT) provided by Xilinx is used to test and debug the GTX transceiver's transmission capabilities and performance. The test result is shown as Fig.7. The bit error rate(BER) is pretty low, and the magnitude is about $10^{-12}$, which indicates that the transceiver link status is normal. We utilized Integrated Logic Analyzer(ILA) core for further testing, which is a logic analyzer that can used to monitor programmable logic signals and ports inside a design project. Test result is shown as Fig.8. The system transmission clock is 50MHz and the internal data bit width is 16bit. System runs at a bandwidth of 1Gbps after 8B / 10B encoding for the internal data. The sender generates a counter within the FPGA in the loopback mode. Furthermore the receiver is able to receive the correct signal.

Fig.7. IBERT test

Fig.8 Test results

## V. CONCLUSION

The DAQ system used for physics experiments is designed and tested. System is able to transmit data stably at the bandwidth of 1Gbps over long distance. In order to achieve better performance, the system will be used for larger bandwidth and longer distance data transmission tests.